\documentclass[%
twocolumn,
superscriptaddress,
nofootinbib,
showpacs,
amsmath,
amssymb,
prl,
aps
]{revtex4-1}

\usepackage{%
 microtype,
 textcomp,
 import,
 graphicx,
 adjustbox,
 dblfloatfix,
 dcolumn,
 natbib,%
 units,
 amsmath,
 nccmath,
 siunitx 
 }
\DeclareSIUnit\gauss{G}
\usepackage[colorlinks=false, pdfborder={0 0 0}]{hyperref}
\usepackage{color}

{\end{pmatrix}\end{medsize}}%

\include{commands}

\graphicspath{{./figures/},{.}}
\begin{document}

\title{Synthetic magnetic fields for cold erbium atoms}

\author{Daniel Babik}
\email{babik@iap.uni-bonn.de}
\affiliation{Institute for Applied Physics, University of Bonn, 53115 Bonn, Germany}

\author{Roberto Roell}
\affiliation{Institute for Applied Physics, University of Bonn, 53115 Bonn, Germany}

\author{David Helten}
\affiliation{Institute for Applied Physics, University of Bonn, 53115 Bonn, Germany}

\author{Michael Fleischhauer}
\affiliation{Department of Physics and Research Center OPTIMAS, University of Kaiserslautern, 67663 Kaiserslautern, Germany}

\author{Martin Weitz}
\affiliation{Institute for Applied Physics, University of Bonn, 53115 Bonn, Germany}

\date{\today}

\begin{abstract}
The implementation of the fractional quantum Hall effect in ultracold atomic quantum gases remains, despite substantial advances in the field, a major challenge. Since atoms are electrically neutral, a key ingredient is the generation of sufficiently strong artificial gauge fields. Here we theoretically investigate the synthetization of such fields for bosonic erbium atoms by phase imprinting with two counterpropagating optical Raman beams. Given the nonvanishing orbital angular momentum of the rare-earth atomic species erbium in the electronic ground state and the availability of narrow-line transitions, heating from photon scattering is expected to be lower than in atomic alkali-metal species. We give a parameter regime for which strong synthetic magnetic fields with good spatial homogeneity are predicted. We also estimate the size of the Laughlin gap expected from the $s$-wave contribution of the interactions for typical experimental parameters of a two-dimensional atomic erbium microcloud. Our analysis shows that cold rare-earth atomic ensembles are highly attractive candidate systems for experimental explorations of the fractional quantum Hall regime.
\end{abstract}


\maketitle

\section{I. Introduction}
To this date the quantum Hall effect is an active frontier of research as it is the hallmark of systems with topolo\-gical order. For two-dimensional (2D) electron gases both the integer and the fractional quantum Hall effect (FQHE) have been observed in strong magnetic fields \cite{Yoshioka2002,Tsui1982}. The key characteristic of a quantum Hall system is the nontrivial topology of its gapped many-body ground state, characterized by a nonvanishing Chern-number. Associated with this are a number of fascinating features such as topologically protected edge states. In the presence of interactions, gapped ground states exists with fractional fillings, and the corresponding FQH states possess a number of
additional interesting properties such as fractional topological charges and anyonic excitations.
Atoms are electrically neutral and for these systems effective magnetic fields must be emulated e.g. using trap rotation \cite{Bretin2004,Schweikhard2004}, lattice shaking \cite{Hauke2012}, or phase imprinting via photon recoil \cite{Lin2009,Aidelsburger2013,Kennedy2013} methods.\par
There are two physical regimes for bosonic atoms in a synthetic magnetic field. If the applied artificial field is small the ground state remains a Bose-Einstein condensate (BEC), 
characterized by a macroscopically occupied single-particle
 wavefunction. The artificial magnetic field induces vortices in the condensate, but as long as the density of vortices is small compared to the density of atoms the condensate is not destroyed. In the ground state the vortices form a regular structure closely related to the Abrikosov lattice of type-II superconductors \cite{Abrikosov1957}. 
 This regime is experimentally well accessible. For much larger values of the applied artificial magnetic field when the density of vortices approaches that of the atomic gas, the vortex lattice melts, and in the presence of interactions the ground state can become a bosonic quantum Hall liquid. Specifically, for a filling factor of $\nu = 1/2$ (which applies when the atom number $N$ equals half the number of vortices, or magnetic flux quanta, $\Phi = q B/h$),  the ground state is a Laughlin state \cite{Cooper2001,Regnault2003}, which is particularly interesting since some of the excitations above this ground state having anyonic character \cite{Paredes2001}.\par
The current experimental challenge lies in generating synthetic gauge fields for atoms that are strong enough to reach the fractional quantum Hall regime. To surpass technical li\-mitations of direct atomic cloud rotation schemes 
one can apply phase imprinting methods via Raman manipulation \cite{Juzeliunas2004,Ruseckas2005,Juzeliunas2006,Zhu2006,Spielman2009} for the generation of gauge fields. Phase imprinting methods at the present stage have used alkali-metal atoms with an $S$-electronic ground state ($L = 0$), for which Raman transitions are only possible between ground state sublevels due to the additional fine- and hyperfine structure, see Fig.~\ref{fig:Twolevelscheme}(a). This impedes the use of laser detunings above the fine structure splitting of the upper electronic state, and leads to a limitation of the possible coherence time. In principle, atoms with a $P$-electronic ground state, see Fig.~\ref{fig:Twolevelscheme}(b) for a corresponding level scheme, would be attractive candidates, however typical atomic species, as the oxygen atom, are technically difficult to laser cool due to technically inconvenient UV electronic transition wavelengths and a large number of required repumping lasers.\par
Cui et al. proposed the use of lanthanide atoms for the generation of gauge fields \cite{Cui2013}.
Specifically they considered the case of dysprosium atoms and a $\sigma^+ - \pi$ optical polarization confi\-guration of Raman beams inducing a $\Delta m = 1$ ground state \mbox{coupling} scheme. Atomic species such as dysprosium, erbium, or thulium fulfill the requirement of an orbital angular momentum $L > 0$ in the electronic ground state. Moreover, for these atomic systems laser cooling and the production of degenerate atomic quantum gases is feasible, see Refs. \cite{Lu2011,Aikawa2012,Aikawa2014,Ulitzsch2017,Kalganova2015} for experimental work realizing both atomic Bose-Einstein condensates or degenerate Fermi gases using corresponding isotopes, or a cold atomic gas for the thulium case. When tuning into the vicinity of suitable electronic transitions, one expects that Raman transitions with far-off-resonant optical beams and cor\-respondingly low heating rates become possible. 
A characteristic measure is the ratio between the upper state linewidth $\Gamma$ and the maximum useable detuning $\Delta$, which is of order of the upper state fine structure splitting.
While for the case of rubidium this ratio is about $\sim 10^{-5}$, a linewidth to detuning ratio $\Gamma / \Delta \simeq 10^{-7}$ seems feasible in rare-earth atomic systems, see also a corresponding estimation in earlier work \cite{Dalibard2015}. Similarly, in these systems one expects that state dependent dipole trapping with long coherence times can be realized. Another, in many cases favorable property of many lanthanides is their large magnetic moment, giving rise to a large dipole-dipole interaction \cite{Kadau2016,Baier2018}.\par
In the present work, we present a scheme for the gene\-ration of artificial magnetic fields for erbium atoms using phase imprinting from two counterpropagating Raman beams with opposite circular polarization, inducing a $\Delta m = 2$ Raman coupling between ground state Zeeman sublevels, in the presence of a transverse gradient of a real magnetic field. The latter is needed to break time-reversal symmetry, a necessary ingredient of the quantum Hall effect. For suitable values of the two-photon Rabi frequency, we obtain a spatially very uniform synthetic magnetic field. We consider some examples for possible submerged shell transitions of the erbium atom to implement the Raman coupling. Finally, using known theory results for the Laughlin gap arising from $s$-wave interactions, we derive estimates for this interaction-induced splitting for a two-dimensional atomic quantum gas subject to the strong synthetic magnetic field.\par
In the following, section II discusses the generation of light-induced magnetic fields in a three-level system, and section III specifies this for some atomic erbium narrow-line transitions. Subsequently, section IV gives an estimation of the size of the Laughlin gap, and section V closes with conclusions.

\section{II. Synthetic magnetic fields for three-level atoms}
\begin{figure}[]
\hbox{\hspace{-0.2em}\includegraphics[width=0.49\textwidth]{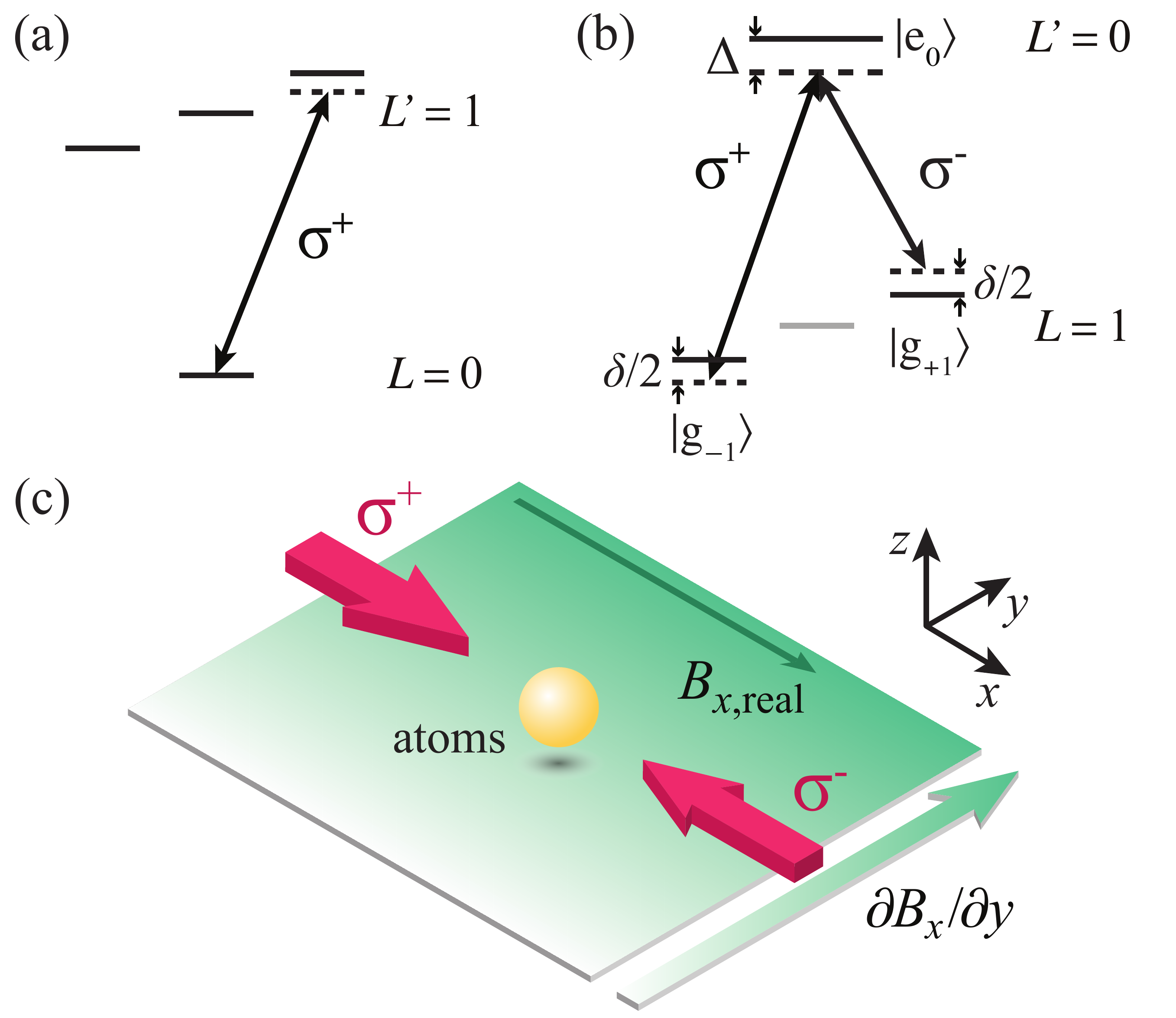}}
\vspace*{-3mm}
\caption{(a) Reduced level scheme of alkali-metal atoms, with an $S$-electronic ground state ($L=0$). (b) Reduced level scheme for a transition from a ground state with $L=1$ to an electronically excited state with $L^\prime=0$. This system gives an example for an electronic transition starting from a higher orbital angular momentum ground state, for which even with radiation far detuned from the electronically excited state Raman transitions between different ground state spin projections become possible (the ground state $|g_0 \rangle$ is shown in gray, because it is not relevant for the atom-light coupling here). For the shown case of $L=1$, the Raman transitions can be driven with a $\sigma^+ - \sigma^-$ optical polarization configuration, inducing Raman transitions with $\Delta m_{\mathrm{F}}=2$. (c) Schematic for synthetization of an artificial magnetic field for atoms using optical driving with two counterpropagating Raman beams and a transverse gradient of the (real) magnetic field.} 
\label{fig:Twolevelscheme}
\end{figure}

For the sake of simplicity of the discussion, we start by considering the generation of a synthetic magnetic field for a three-level atom with two stable ground state levels $|g_{+1}\rangle$ and $|g_{-1}\rangle$ and one spontaneously decaying excited state level $|e_{0}\rangle$, as shown in Fig.~\ref{fig:Twolevelscheme}(b). The index 
denotes the corresponding Zeeman quantum number. The suggested implementation follows the work of Spielman \cite{Spielman2009} developed for alkali-metal atoms, however we here consider the atom to be driven by two far-detuned counterpropagating laser beams in a $\sigma^+ - \sigma^-$ polarization configuration, which results in Raman coupling between ground state sublevels $|g_{+1}\rangle$ and $|g_{-1}\rangle$ with $\Delta m = 2$. The general idea is to construct a Hamiltonian offering an atomic dispersion that mimics that of a charged particle in the presence of a position-dependent vector potential $\vec{A}^\ast$, so that a synthetic magnetic field $\vec{B}^\ast = \nabla \times \vec{A}^\ast$ emerges. 
(The superscript $*$ denotes the artificial, effective magnetic field and is used in order to distinguish it from a real magnetic field.)
This can be achieved with a transversal gradient of the (real) magnetic field, which leads to a two-photon detuning $\delta = \omega_+ - \omega_- - \omega_{\mathrm{Z}}$, where $\omega_+$ and $\omega_-$ denote the laser frequencies with corresponding polarizations and $\hbar \omega_{\mathrm{Z}}$ is the energetic difference between $|g_{+1}\rangle$ and $|g_{-1}\rangle$ that is position-dependent. Assume both the magnetic field $\vec{B} = B_x \vec{e}_x$ and the counterpropagating laser beams oriented along the $x$-axis and a magnetic field gradient along the $y$-axis, see also Fig.~\ref{fig:Twolevelscheme}(c). We now have $B_x(y) = B_{0,x} + y \partial B_x / \partial y$, which realizes a position-dependent Raman detuning \mbox{$\delta(y) = 2 g \mu_{\mathrm{B}} y \partial B_x / \partial y$}, where $g$ is the atomic Land\'{e} $g$-factor and $\mu_{\mathrm{B}}$ the Bohr magneton.

\begin{figure}[]
\hbox{\hspace{-0.66em}\includegraphics[width=0.5\textwidth]{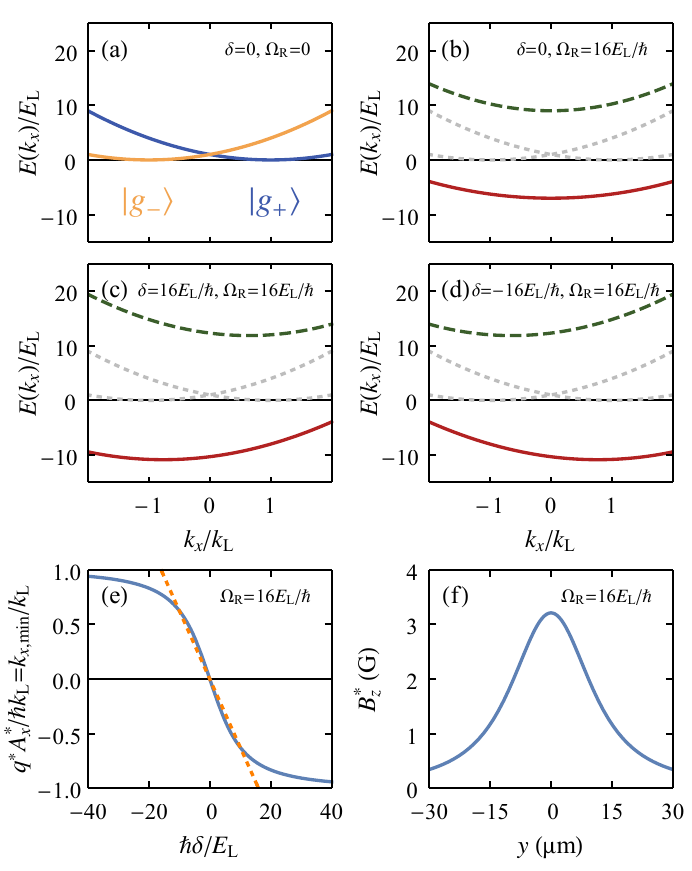}}
\vspace*{-3mm}
\caption{Synthetization of magnetic fields in a three level configuration. (a),(b) Energy quasimomentum dispersion relation for undressed ($\Omega_{\mathrm{R}} = 0$) and a dressed case (\mbox{$\Omega_{\mathrm{R}} = 16 E_{\mathrm{L}}/\hbar$}), both for a vanishing two-photon detuning $\delta$. (c),(d) Corresponding curves for the nonvanishing detuning values of $\delta = \pm 16 E_{\mathrm{L}}/\hbar$ respectively for $\Omega_{\mathrm{R}} = 16 E_{\mathrm{L}}/\hbar$. The energy curves for the undressed case are plotted in (b)-(d) as gray dotted lines for comparison. (e) Generated vector potential $q^\ast A^\ast_x(\delta) / \hbar$ versus the two-photon detuning $\delta$ for \mbox{$\Omega_{\mathrm{R}} = 16 E_{\mathrm{L}}/\hbar$} (blue solid line) and the dependence obtained from a Taylor expansion up to lowest order in $\delta$ ($k_{x,\mathrm{min}}^{\mathrm{Taylor}} / k_{\mathrm{L}} = - \delta / \Omega_{\mathrm{R}}$, see text), yielding a linear slope (orange dotted line). (f) The generated synthetic magnetic field, when applying a transverse detuning gradient $\delta^\prime / (2 \pi) = 2.66 \, \unit{kHz/\mu m}$ with a gradient of the real magnetic field versus position $y$. Here $q^\ast = e$ was assumed.}
\label{fig:Rubidium}
\end{figure}

For the counterpropagating laser beam configuration $\sim 2 \hbar k_{\mathrm{L}}$ momentum per Raman transition is transferred to the atoms, with $\vec{k}_{\mathrm{L}} = k_L \vec{e}_x$ and $k_{\mathrm{L}} = 2 \pi / \lambda$. In the following, $|g_{\alpha},\vec{p}\rangle$ denotes an atom in the internal state $g_{\alpha}$ and momentum $\vec{p}$. The resulting effective Hamiltonian for a single atom confined to the $x$-$y$ plane can be written in the basis of the coupled levels \mbox{$|g_{-1},\hbar (\vec{k}+\vec{k}_{\mathrm{L}})\rangle$} and \mbox{$|g_{+1},\hbar (\vec{k}-\vec{k}_{\mathrm{L}})\rangle$} as

\begin{eqnarray}
\resizebox{0.9\hsize}{!}{$
\hat{H} = \left(\begin{matrix}
\frac{\hbar^2 (k_x - k_{\mathrm{L}})^2}{2 m}+\frac{\hbar \delta(y)}{2} & \frac{\hbar \Omega_{\mathrm{R}}}{2}\\[6pt]
\frac{\hbar \Omega_{\mathrm{R}}}{2} & \frac{\hbar^2 (k_x + k_{\mathrm{L}})^2}{2 m}-\frac{\hbar \delta(y)}{2}
\end{matrix} \right) + \frac{\hbar^2 k_y^2}{2 m} , \label{eq:Hmatrix}
$}
\end{eqnarray}

\noindent where $\Omega_{\mathrm{R}}$ denotes the effective Rabi frequency of the two-photon Raman transition. Fig.~\ref{fig:Rubidium}(a) shows the variation of the eigenstates of the uncoupled system (i.e. for $\Omega_{\mathrm{R}} = 0$), for which we obtain the usual parabolic dispersion centered at $-k_{\mathrm{L}}$ and $k_{\mathrm{L}}$ for states $|g_{-1}\rangle$ and $|g_{+1}\rangle$ respectively. Fig.~\ref{fig:Rubidium}(b) shows the  dispersion for a nonvanishing Raman coupling (\mbox{$\Omega_{\mathrm{R}} = 16 E_{\mathrm{L}}/\hbar$}, where \mbox{$E_{\mathrm{L}} = \hbar^2 k^2_{\mathrm{L}}/2 m$} denotes the recoil energy), resulting in a dressing of the energy levels. In general one finds that  in the presence of the dressing for $\Omega_{\mathrm{R}} \gtrsim 4 E_{\mathrm{L}}/\hbar$ the two resulting energy curves have a combined single minimum, which for $\delta = 0$ appears at $k_x = 0$, but can be shifted from that position in $k$-space by a nonvanishing value of $\delta$.

We are interested in the following in the lower of the two dressed energy levels, with the dispersion shown as a solid red line in Fig.~\ref{fig:Rubidium}(b)-(d)  , whose position of the minimum depends on the value of the Raman detuning, see Figs.~\ref{fig:Rubidium}(c),(d). In the presence of the gradient of the real magnetic field, this Raman detuning depends in turn on the transverse position $y$, see also Fig.~\ref{fig:Rubidium}(e). The effective Hamiltonian for the lower dressed state can thus be approximated as

\begin{subequations}\label{eq:Hcharged}
\begin{align}
\hat{H}_{\mathrm{eff}} & \approx E_0 + \frac{\hbar^2 (k_x - k_{\mathrm{x,min}}(y))^2}{2 m^\ast} + \frac{\hbar^2 k_y^2}{2 m}\label{eq:HchargedpartA}\\
& = E_0 + \frac{\hbar^2}{2 m^\ast} \left( k_x - \frac{q^\ast A_x^\ast(y)}{\hbar}\right)^2 + \frac{\hbar^2 k_y^2}{2 m} , \label{eq:HchargedpartB}
\end{align}
\end{subequations}

\noindent where $k_{\mathrm{x,min}}(y)$ denotes the wavevector at which the described minimum of the dispersion curve occurs, and $m^\ast$ denotes an effective mass for the motion along the $x$-direction. In the second equation (Eq.~\ref{eq:HchargedpartB}) we have used the replacement $k_{\mathrm{x,min}}(y) = q^\ast A_x^\ast(y)/\hbar$, where $A_x^\ast$ is the synthetic vector potential discussed above and $q^\ast$ is a synthetic charge, which will be chosen by convenience. Note that both the effective mass $m^\ast$ and $k_{\mathrm{x,min}}$ (correspondingly also the synthetic vector potential $A_x^\ast$ and the synthetic magnetic field $B_z^\ast$, the latter as introduced below) depend on the used value of the effective Rabi frequency $\Omega_{\mathrm{R}}$.

Given the transverse detuning variation from the gradient of the (real) magnetic field, we readily expect a nonvanishing value of the synthesized magnetic field along the $z$-axis: \mbox{$B_z^\ast = - \partial A_x^\ast(y) / \partial y = - \hbar/q^\ast \partial k_{\mathrm{x,min}}(y) / \partial y$}, with \mbox{$\delta^\prime = \partial \delta / \partial y = 2g\mu_{\mathrm{B}} \partial B_x / \partial y$} as the detuning gradient. We arrive at

\begin{eqnarray}
B_z^\ast = - \frac{\hbar \delta^\prime}{q^\ast} \frac{\partial k_{x,\mathrm{min}}(y)}{\partial \delta}. \label{eq:BzFinal}
\end{eqnarray}

Near $y = 0$, for which $\delta \approx 0$, the synthetic vector potential $A_x^\ast$ varies linearly with the two-photon detuning $\delta$, and correspondingly the transverse position $y$ (Fig.~\ref{fig:Rubidium}(e)). The generated synthetic magnetic field has a maximum at $y = \delta(y) = 0$. The magnitude of the synthetic field in the center can for $\Omega_{\mathrm{R}} \gg E_{\mathrm{L}}/\hbar$ be estimated when noting that in the limit of a detuning $\delta \gtrsim \Omega_{\mathrm{R}}$ ($\delta \lesssim - \Omega_{\mathrm{R}}$) we have $k_{x,\mathrm{min}} = - k_{\mathrm{L}} (+ k_{\mathrm{L}})$ respectively (compare also Fig.~\ref{fig:Rubidium}(b)-(d)), so that one expects a slope near $\delta = 0$ of order $\partial k_{x,\mathrm{min}} / \partial \delta \approx - k_{\mathrm{L}} / \Omega_{\mathrm{R}}$, from which we find $B_z^\ast(y=0) \approx \hbar k_{\mathrm{L}} \delta^\prime / q^\ast \Omega_{\mathrm{R}}$. These results are also obtained from a Taylor expansion of the analytically obtained expression of the position of the minimum up to lowest order in $\delta$ (for again $\Omega_{\mathrm{R}} \gg E_{\mathrm{L}}/\hbar$). The orange dashed line in Fig.~\ref{fig:Rubidium}(e) shows the based on this expansion derived value of the synthetic vector potential versus the detuning.

When numerically determining the minimum of the low energy dispersion for corresponding parameters, we arrive at the "exact" dependence, as shown by the blue line in Fig.~\ref{fig:Rubidium}(e). Here again $\Omega_{\mathrm{R}} = 16 E_{\mathrm{L}}/\hbar$ was used. Fig.~\ref{fig:Rubidium}(f) shows the corresponding spatial variation of the synthetic magnetic field versus the position along the $y$-axis (as derived using Eq.~\ref{eq:BzFinal}). The assumed experimental parameters for the magnetic field gradient, and also the obtained magnitude and spatial variation of the synthetic magnetic field are comparable to the case of the rubidium experiment of \cite{Spielman2009}. The different transferred momentum of the Raman transitions with counterpropagating laser beams introduces modifications \mbox{of order below} a factor of 2. A clear disadvantage of the three-level scheme is the inhomogeneity of the effective magnetic field $B^*_z$, as shown in Fig.~\ref{fig:Rubidium}(f).

\section{III. Synthetic magnetic fields for the atomic erbium case}
We now discuss the possibility to generate synthetic magnetic fields for a specific rare-earth system, atomic erbium, for the transitions $4f^{12}6s^2({}^3\textrm{H}_6) \rightarrow 4f^{11}({}^4\textrm{I}^0_{15/2})5d_{5/2}6s^2(15/2, 5/2)_{J^\prime}^0$, with $J'=5,6$ and $7$ at transition wavelengths of $\lambda = 877 \, \mathrm{nm}$, $\lambda = 847 \, \mathrm{nm}$, and $\lambda = 841 \, \mathrm{nm}$ respectively. These transitions excite an electron within the incompletely filled ``submerged'' $f$-shell of the atom and all have a relatively small natural linewidth, e.g. $\Gamma/2 \pi = 8.0 \, \mathrm{kHz}$ for the \mbox{$J=6 \rightarrow J'=7$} transition near $\lambda = 841 \, \mathrm{nm}$ wavelength \cite{Ban2005}. Given the comparatively large energy difference to neighbouring levels in terms of the linewidths, the systems are very attractive for Raman manipulation with far detuned optical beams. Ultimately, we expect the atomic lifetime to be limited by off-resonant scattering from e.g. the strong (\mbox{$\Gamma_{\mathrm{blue}}/2 \pi \approx 28 \, \mathrm{MHz}$}) blue cooling transition near $401 \, \mathrm{nm}$, which is detuned by an amount of order of the optical frequency. This sets a limit on the usable detuning from the upper state from the narrow-line transition of order $\Delta / \Gamma \simeq 10^7$, and within this limit we assume in the following that off-resonant contributions from other excited states can be neglected.

\begin{figure}
\hbox{\hspace{-0.66em}\includegraphics[width=0.5\textwidth]{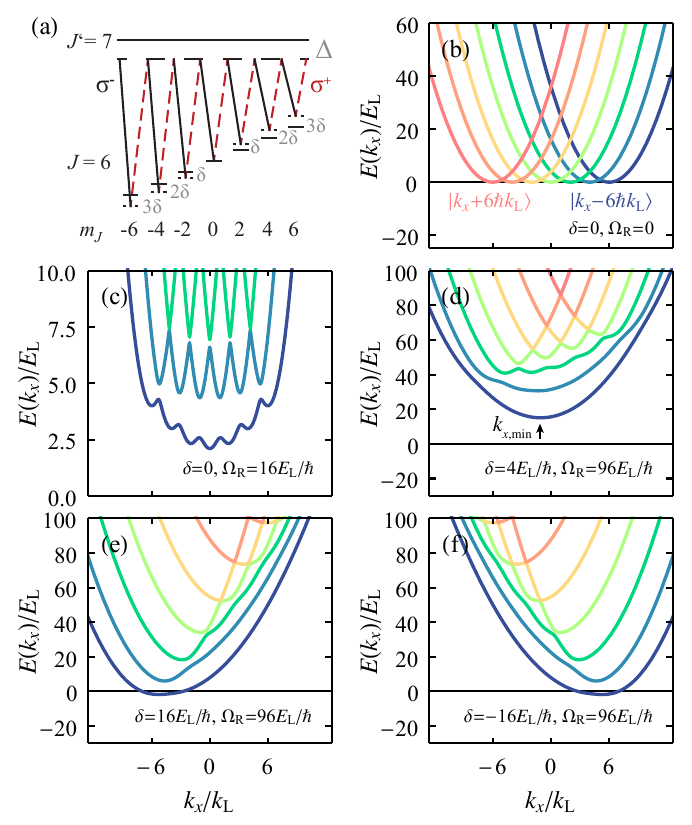}}
\vspace*{-3mm}
\caption{(a) Relevant atomic erbium levels for the \mbox{$J = 6 \rightarrow J' = 7$} transition driven by Raman beams in a $\sigma^+ - \sigma^-$ optical polarization configuration. The Raman beams are irradiated in a counterpropagating geometry. (b) Dispersion relation $E(k_x)$ of the seven undressed (with $\Omega_{\mathrm{R}} = 0$) states for $\delta = 0$. (c) Dispersion of the dressed state system with moderate Raman coupling (\mbox{$\Omega_{\mathrm{R}} = 8 E_{\mathrm{L}}/\hbar$}, for which \mbox{$\Omega_{\mathrm{R}} < m_{F \mathrm{,max}}^2 E_{\mathrm{L}}/\hbar$}) and $\delta = 0$. (d) Dispersion for a larger value of the Raman coupling (\mbox{$\Omega_{\mathrm{R}} = 96 E_{\mathrm{L}}/\hbar$}, for which \mbox{$\Omega_{\mathrm{R}} > m_{F \mathrm{,max}}^2 E_{\mathrm{L}}/\hbar$}, with \mbox{$m_{F \mathrm{,max}} = 6$}), for which the lowest energetic dressed state level has a near parabolic shape. Here a nonvanishing two-photon detuning $\delta=4 E_{\mathrm{L}}/\hbar$ was used, resulting in a minimum of the dispersion curve at $k_{x \mathrm{,min}} \neq 0$. (e),(f) Corresponsing curves for the larger positive/negative detuning values $\delta = \pm 16 E_{\mathrm{L}}/\hbar$ respectively.}
\label{fig:Erbiumscheme}
\end{figure}

Despite of the small scattering rate for radiation correspondingly tuned in the vicinity of such an inner-shell transition, scalar, vector, and tensor polarizabi\-lities become comparable \cite{Lepers2014}. We assume a nuclear spin of $I = 0$, as is the case for all stable bosonic erbium isotopes (e.g. ${}^{168}\mathrm{Er}$), so that $F = J$. As the ${}^3\textrm{H}_6$ ground state of atomic erbium possesses a total angular momentum of $J = 6$ (with $L = 5, S = 1$), 13 $m_F$-sublevels exist. Our  Raman coupling scheme uses a $\sigma^+-\sigma^-$ configuration, coupling only states with $\Delta m_F = \pm 2$, so that 7 ground-state sublevels $|g_{\alpha}\rangle$, with $m_F = \alpha$ and \mbox{$\alpha = -6, -4, ... , 6$}, are coupled by the Raman beams, see Fig.~\ref{fig:Erbiumscheme}(a) for the coupling scheme of the \mbox{$J=6 \rightarrow J'=7$} transition.

The laser electric field is $\vec{E} = E_{0,+} \vec{e}_+ \cos(k_{\mathrm{L}} x - \omega_+ t) + E_{0,-} \vec{e}_- \cos(- k_{\mathrm{L}} x - \omega_- t)$, where $E_{0,\pm}$ denotes the field amplitudes of the $\sigma^+$, $\sigma^-$ polarized optical beams and $\vec{e}_\pm$ are the corresponding unit polarization vectors.

The relative strength of the coupling between a certain ground state sublevel $|g_{\alpha}\rangle$ component and an excited state component $|e_{n}\rangle$ with $\alpha = n \pm 1$ is characterized by the corresponding Clebsch-Gordan coefficient $c_{\alpha,n}$. For a list of Clebsch-Gordan coefficients for the three erbium transitions mentioned above, see the appendix. The laser coupling between levels can be written in the form \mbox{$\Omega_{\pm} c_{\alpha,\alpha \pm 1} = \langle e_{\alpha \pm 1} | \vec{e}_\pm \vec{d} | g_\alpha \rangle E_{0,\pm}/\hbar$}, where $\vec{d}$ denotes the dipole operator, and $\Omega_+$,$\Omega_-$ the Rabi frequencies for the $\sigma^+$,$\sigma^-$ polarized waves respectively for a transition with a Clebsch-Gordan coefficient of unity. For a large detuning $\Delta$ from the excited levels, the upper states can be adiabatically eliminated, and we arrive at an effective interaction Hamiltonian for the coupling to the laser fields

\begin{eqnarray}
\begin{split}
\hat{H}_{\mathrm{eff}}^\prime & = \frac{p^2}{2 m} + \sum_{\substack{\alpha = -6 \\ \alpha / 2 \in \mathbb{Z}}}^{6} \hbar \left[ \omega_{\mathrm{AC},\alpha} -\frac{\alpha}{2} \delta \right] |g_\alpha \rangle \langle g_\alpha |\\
& + \sum_{\substack{\alpha = -6 \\ \alpha / 2 \in \mathbb{Z}}}^{4} \frac{\hbar \tilde{\Omega}_{\mathrm{R},\alpha,\alpha + 2}}{2} |g_\alpha \rangle \langle g_{\alpha + 2} | \mathrm{e}^{-i2k_{\mathrm{L}}x}\\
& + \sum_{\substack{\alpha = -4 \\ \alpha / 2 \in \mathbb{Z}}}^{6} \frac{\hbar \tilde{\Omega}_{\mathrm{R},\alpha,\alpha - 2}}{2} |g_\alpha \rangle \langle g_{\alpha - 2} | \mathrm{e}^{i2k_{\mathrm{L}}x} , \label{eq:HeffRaman}
\end{split}
\end{eqnarray}

\noindent where $\tilde{\Omega}_{\mathrm{R},\alpha,\alpha \pm 2} = c_{\alpha,\alpha \pm 1}c_{\alpha \pm 2,\alpha \pm 1} \Omega_{\pm} \Omega_{\mp}/(2 \Delta)$ denote effective two-photon Rabi frequencies between ground state sublevels and $\omega_{\mathrm{AC},\alpha} = (c^2_{\alpha,\alpha + 1} \Omega^2_+ + c^2_{\alpha,\alpha - 1} \Omega^2_-)/(2 \Delta)$ is the AC Stark shift of the ground state sublevels. As we choose a large detuning $\Delta$, any excited-state shifts $\delta_n$ can be neglected. In the basis of eigenstates $|g_{\alpha},\vec{p}+\alpha \hbar \vec{k}_{\mathrm{L}} \rangle$ with $\alpha = -6, -4, ... , 6$, where $\vec{p} = \hbar \vec{k}$, Eq.~\ref{eq:HeffRaman} can be written more explicitely using the matrix form

\begin{figure}[]
\hbox{\hspace{-0.66em}\includegraphics[width=0.5\textwidth]{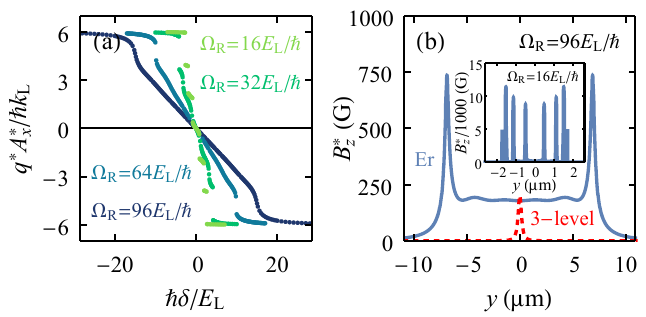}}
\vspace*{-3mm}
\caption{(a) Variation of the synthetic vector potential $A^\ast_x$ versus the two-photon detuning $\delta$ for different values of the (unity Clebsch-Gordan coefficients) effective two-photon Rabi frequency $\Omega_{\mathrm{R}}$. For too low values of $\Omega_{\mathrm{R}}$ (roughly below \mbox{$m_{F \mathrm{,max}}^2 E_{\mathrm{L}}/\hbar$}, with \mbox{$m_{F \mathrm{,max}} = 6$}), no continuous variation is observed. (b) Corresponding synthetic magnetic field (blue) versus position along the $y$-axis for a transverse gradient of the real magnetic field of \mbox{$\partial B_x / \partial y = 70.3 \, \unit{G/cm}$}. While the synthetic field is spikey for the case of small values of $\Omega_{\mathrm{R}}$ (see inset for $\Omega_{\mathrm{R}} = 16 E_{\mathrm{L}}/\hbar$), for \mbox{$\Omega_{\mathrm{R}} = 96 E_{\mathrm{L}}/\hbar$} a over a relatively large spatial region good spatial homogeneity is reached (see main panel). For comparison, also the spatial variation of the synthetic magnetic field obtained for a pure three level system as shown in Fig.~\ref{fig:Twolevelscheme}(b) is shown (red dashed line), where a two-photon Rabi frequency $\Omega_{\mathrm{R}} = 16 E_{\mathrm{L}}/\hbar$, $g = 1$, and a magnetic field gradient of $595 \, \unit{G/cm}$ was assumed, for which the desired value of $\hbar \omega_{\mathrm{c}} = E_{\mathrm{L}}$ in the center (at $y = 0$) is achieved.}
\label{fig:VecpotSMF}
\end{figure}

\begin{widetext}
\begin{eqnarray}
\hat{H}_{\mathrm{eff}} = \left( \begin{matrix} H_{-6,-6} & \tilde{\Omega}_{\mathrm{-6,-4}} & 0 & 0 & 0 & 0 & 0\\
\tilde{\Omega}_{\mathrm{-4,-6}} & H_{-4,-4} & \tilde{\Omega}_{\mathrm{-4,-2}} & 0 & 0 & 0 & 0\\
0 & \tilde{\Omega}_{\mathrm{-2,-4}} & H_{-2,-2} & \tilde{\Omega}_{\mathrm{-2,0}} & 0 & 0 & 0\\
0 & 0 & \tilde{\Omega}_{\mathrm{0,-2}} & H_{0,0} & \tilde{\Omega}_{\mathrm{0,2}} & 0 & 0\\
0 & 0 & 0 & \tilde{\Omega}_{\mathrm{2,0}} & H_{2,2} & \tilde{\Omega}_{\mathrm{2,4}} & 0\\
0 & 0 & 0 & 0 & \tilde{\Omega}_{\mathrm{4,2}} & H_{4,4} & \tilde{\Omega}_{\mathrm{4,6}}\\
0 & 0 & 0 & 0 & 0 & \tilde{\Omega}_{\mathrm{6,4}} & H_{6,6}
\end{matrix} \right) ,
\end{eqnarray}
\end{widetext}

\noindent where $H_{\alpha , \alpha} = \hbar (\omega_{\mathrm{AC,\alpha}} - \alpha \delta / 2)+ \hbar^2 ((k_x + \alpha k_L)^2 +k_y^2)/ 2 m$ and $\tilde{\Omega}_{\alpha, \alpha \pm 2} = \hbar \tilde{\Omega}_{\mathrm{R,\alpha,\alpha \pm 2}} / 2$. To find the eigenenergies of the multi-level system, we numerically solve the eigensystem $(4)$. Fig.~\ref{fig:Erbiumscheme}(b) shows the energy dispersion curves for $\delta = 0$ of the uncoupled system, and Figs.~\ref{fig:Erbiumscheme}(c) and ~\ref{fig:Erbiumscheme}(d) for different values of $\Omega_{\mathrm{R}} = 8 E_{\mathrm{L}}/\hbar$ and $96 E_{\mathrm{L}}/\hbar$ respectively, where \mbox{$\Omega_{\mathrm{R}} = \Omega_{\pm} \Omega_{\mp}/(2 \Delta)$} denotes the effective two-photon Rabi frequency for Clebsch-Gordan coefficients of unity. We here are interested in the dispersion of the lowest energetic eigenstate. While for the lower value of the two-photon Rabi coupling (Fig.~\ref{fig:Erbiumscheme}(c)) the curve has multiple minima, the plot shown in Fig.~\ref{fig:Erbiumscheme}(d) with $\Omega_{\mathrm{R}} = 96 E_{\mathrm{L}}/\hbar$ depicts a smooth, near parabolic dispersion of the low energy dressed state. More generally, for the $J = 6 \rightarrow J^\prime = 7$ transition we find that for Rabi frequencies $\hbar \Omega_{\mathrm{R}} \gtrsim (m_{F \mathrm{,max}} \hbar k_{\mathrm{L}})^2/(2 m) = m^2_{F \mathrm{,max}} E_{\mathrm{L}}$, with $m_{F \mathrm{,max}} = 6$, corresponding to the recoil energy associated with the momentum difference between atoms in an outermost and a central Zeeman sublevel, the dispersion can be approximated as $E(\delta) = E_0 + \hbar(k_x - k_{x\mathrm{,min}}(\delta))^2/(2 m^\ast)$ for not too large values of the detuning $\delta$. With the identification $A_x^\ast (\delta) = \hbar k_{x\mathrm{,min}}(\delta)/q^\ast$ and noting that in the presence of the transverse gradient of the real magnetic field $\delta = \delta(y)$, we find that one can describe the atomic dynamics also in the multi-level case by an effective Hamiltonian of the form of Eq.~\ref{eq:HchargedpartB}. In addition, a scalar potential emerges. Figs.~\ref{fig:Erbiumscheme}(e) and (f) give the corresponding dispersion curves for detuning values of $\delta = 16 E_{\mathrm{L}}/\hbar$ and $-16 E_{\mathrm{L}}/\hbar$ respectively (with again $\Omega_{\mathrm{R}} = 96 E_{\mathrm{L}}/\hbar$). The blue line in Fig.~\ref{fig:VecpotSMF}(a) shows the dependence of the generated synthetic vector potential versus $\delta$, which varies smoothly between $-6\hbar k_{\mathrm{L}}/q^\ast$ and $6\hbar k_{\mathrm{L}}/q^\ast$. Fig.~\ref{fig:VecpotSMF}(b) shows the corresponding synthetic magnetic field for a detuning gradient of $\delta^\prime/(2 \pi) = 21 \, \unit{kHz/\mu m}$, as obtained with a gradient of the real magnetic field of $70.3 \, \unit{G/cm}$ for the erbium case with $g = 1.166$. Importantly, the synthetic magnetic field is spatially very uniform over a relatively large distance ($\sim 10 \, \unit{\mu m}$, see below for further discussion), with additional peaks at the edge. For smaller values of the two-photon Rabi frequencies the synthetic magnetic field loses spatial homogeneity, and for values below the multi-photon recoil even becomes spikey (see inset of Fig.~\ref{fig:VecpotSMF}(b)), as understood from the multiple minima of the dispersion curve in this parameter regime (see e.g. Fig.~\ref{fig:Erbiumscheme}(c)). It is interesting to note that also for the three-level scheme, as discussed above in section II, at low values of the Rabi frequencies (for $\Omega_{\mathrm{R}} \lesssim 4 E_{\mathrm{L}}/\hbar$) not a single, but rather two minima appear in the dispersion relation, which similarly as discussed here results in a spiky behavior of the synthetic magnetic field.

\renewcommand{\arraystretch}{1.25} 
\setlength{\tabcolsep}{5.5pt}      
\begin{table}
\begin{adjustbox}{width=\columnwidth,center}
    \begin{tabular}{|c|c|c|c|c|c|c|}
    \hline
    &\multicolumn{3}{c|}{$J' = 5$}& \multicolumn{3}{c|}{$J' = 7$}\\ \hline
    $\Omega_{\mathrm{R}} / E_{\mathrm{L}}$ & $\delta^\prime / 2 \pi$ & $\partial B_x / \partial y$ & $I$ & $\delta^\prime / 2 \pi$ & $\partial B_x / \partial y$ & $I$ \\ \hline
    32 & 9.18 & 28.14 & 4.52 & 10.42 & 31.94 & 4.90\\ \hline
    64 & 15.78 & 48.36 & 8.98 & 14.98 & 45.92 & 9.74\\ \hline
    96 & 24.11 & 73.88 & 13.51 & 22.95 & 70.34 & 14.64\\ \hline

    \end{tabular}
\end{adjustbox}
\caption{Detuning gradient $\delta^\prime / 2 \pi$ (in $\unit{kHz/ \mu m}$), real magnetic field gradient $\partial B_x / \partial y$ (in $\unit{G/cm}$) and Raman beam intensity $I$ (in $\unit{W/mm^2}$) for which a cyclotron frequency of $\hbar \omega_{\mathrm{c}} = E_{\mathrm{L}}$ is reached for the two feasible Raman transitions for different values of the two-photon Rabi frequency $\Omega_{\mathrm{R}}$ for unity Clebsch-Gordan coefficients. \label{tab:Gradients}}
 \end{table}

The value for the gradient of the real magnetic field (\mbox{$\partial B_x / \partial y = 70.3 \, \unit{G/cm}$}) for the erbium $J=6 \rightarrow J^\prime=7$ transition was chosen to reach a ratio of $\hbar \omega_{\mathrm{c}} / E_{\mathrm{L}} = 1$ (with $E_{\mathrm{L}}/(2 \pi \hbar) \simeq 1.68 \, \unit{kHz}$) in the center, where $\omega_c = e B^\ast / m^\ast$ denotes the value of the cyclotron frequency.
According to earlier work (\cite{Dalibard2015}) this is a desirable parameter regime for the observation of fractional quantum Hall physics in such systems. For the magnetic length we find $\ell_{\mathrm{mag}} = \sqrt{\hbar/m \omega_{\mathrm{c}}} \approx 0.19 \, \unit{\mu m}$, yielding an area of order $\mathcal{A} \sim 2 \pi \ell^2_{\mathrm{mag}}$ per flux quantum, or an atomic area density $n \simeq 1/(4 \pi \ell^2_{\mathrm{mag}}) \approx 2 \, \unit{\mu m}^{-2}$ at half filling. The area of spatial homogeneity shown in Fig.~\ref{fig:VecpotSMF}(b) of $\approx 10 \, \unit{\mu m}$ diameter in a circular 2D geometry should thus be sufficient to load up to $\approx 200$ atoms into a Laughlin state. The red dashed line in Fig.~\ref{fig:VecpotSMF}(b) for comparison gives the spatial variation of the synthetic magnetic field for the case of an idealized three-level system (Fig.~\ref{fig:Twolevelscheme}(b)), with parameters as to also obtain $\hbar \omega_{\mathrm{c}} = E_{\mathrm{L}}$ at $y = 0$. Note that typical area densities of cold atom systems differ from values used in electron fractional quantum Hall systems \cite{Yoshioka2002}, so also required (synthetic or real respectively) magnetic field strengths differ.

A ratio $\Delta / \Gamma = 10^7$ is achieved for the \mbox{$J=6 \rightarrow J'=7$} atomic erbium transition for a detuning $\Delta / (2 \pi) \simeq 80 \, \unit{GHz}$. Both the required Raman beam intensity of $\sim 14.6 \, \unit{W/mm^2}$, corresponding to e.g. $\simeq 115 \, \unit{mW}$ beam power on a $100 \, \unit{\mu m}$ beam diameter, and the described value of the magnetic field gradient are experimentally well achievable. For the quoted parameters we have $\tilde{\Omega}_{\mathrm{R},0,\pm 2} = c_{0,\pm 1}c_{\pm 2,\pm 1}\Omega_{\mathrm{R}} = \sqrt{2/13} \cdot 3 \sqrt{5/96} \cdot 96 E_{\mathrm{L}}/\hbar \simeq 26 E_{\mathrm{L}}/\hbar$, which is roughly about a factor two above the value investigated for rubidium in \cite{Spielman2009}. On the other hand, the Clebsch-Gordan coefficients for the $\sigma^+ - \sigma^-$ polarization configuration considered here are more favorable than for the $\sigma^+ - \pi$ case investigated in the rubidium works, so one may expect the ratio of Rabi coupling and spontaneous scattering at comparable detuning for the erbium and rubidium cases to be roughly comparable. As noted above, the lanthanide case is expected to allow for larger values of $\Delta / \Gamma$, which should reduce the influence of spontaneous scattering. Given that for a smooth variation of the low energy dispersion curve with a single minimum $\Omega_{\mathrm{R}}$ should be above $\sim m_{F \mathrm{,max}}^2 E_{\mathrm{L}}/\hbar$, from the point of a low spontaneous scattering rare-earth atoms with not too high values of $m_{F \mathrm{,max}}$ seem advantageous, although this limits the magnitude of the achievable synthetic magnetic flux.

Rare-earth atoms with not too high value of $m_{F \mathrm{,max}}$ also have a reduced magnetic dipole-dipole interaction. This effect is already relevant when comparing the erbium (${}^{168}_{}\mathrm{Er}$) and dysprosium (${}^{164}_{}\mathrm{Dy}$) cases, with ratios of the dipole-dipole interaction and $s$-wave interaction, assuming the background scattering length, of $\epsilon_{\mathrm{dd,}{}^{168}_{}\mathrm{Er}} \simeq 0.4$ and $\epsilon_{\mathrm{dd,}{}^{164}_{}\mathrm{Dy}} \simeq 1.45$ respectively. We are aware that dipolar physics still is important also for the erbium case \cite{Martin2017}.  

In general, the magnitude of the synthetic field can, as understood from Eq.~\ref{eq:BzFinal}, also for the multilevel case be varied by choice of a suitable detuning gradient $\delta^\prime$, as tuneable experimentally via the gradient of the real magnetic field $\partial B_x / \partial y$. In a related manner as discussed for the three-level case, for the here considered erbium transition we, assuming $\Omega_{\mathrm{R}} > m_{F,\mathrm{max}}^2 E_{\mathrm{L}}/\hbar$ in the large detuning limit of $m_{F,\mathrm{max}} \delta \gtrsim \Omega_{\mathrm{R}}$ ($m_{F,\mathrm{max}} \delta \lesssim - \Omega_{\mathrm{R}}$), arrive at $k_{x,\mathrm{min}} = - m_{F,\mathrm{max}} k_{\mathrm{L}}$ ($m_{F,\mathrm{max}} k_{\mathrm{L}}$) respectively, see also Figs.~\ref{fig:Erbiumscheme}(e),(f). The synthetic field in the central spatial region (near $y = 0$) will be of order $B_z^\ast \sim \hbar k_{\mathrm{L}} m_{F,\mathrm{max}}^2 \delta^\prime / (\Omega_{\mathrm{R}} q^\ast)$ for the here relevant case of $\Omega_{\mathrm{R}} > m_{F,\mathrm{max}}^2 E_{\mathrm{L}} / \hbar$. The dependence of the synthetic field on the effective Rabi frequency $\Omega_{\mathrm{R}}$ is understood from the influence of the coupling on the dressed system dispersion relation, see also the numerically obtained plots of Fig.~\ref{fig:VecpotSMF}(a). Given the required smoothness of the dispersion curve $\Omega_{\mathrm{R}}$ is not truly a free parameter, but should rather be choosen as a few times $m_{F,\mathrm{max}}^2 E_{\mathrm{L}} / \hbar$. Correspondingly, the dependence of the synthetic field on the maximum Zeeman quantum number $m_{F,\mathrm{max}}$ effectively cancels, given the for large values of $m_{F,\mathrm{max}}$ required increased Rabi coupling. On the other hand, the for large values of $m_{F,\mathrm{max}}$ increased possible maximum momentum transfer $m_{F,\mathrm{max}} \hbar k_{\mathrm{L}}$ translates into a larger spatial area over which the synthetic magnetic field is imprinted, and correspondingly a higher synthetic flux.


\begin{figure}[]
\hbox{\hspace{-0.33em}\includegraphics[width=0.5\textwidth]{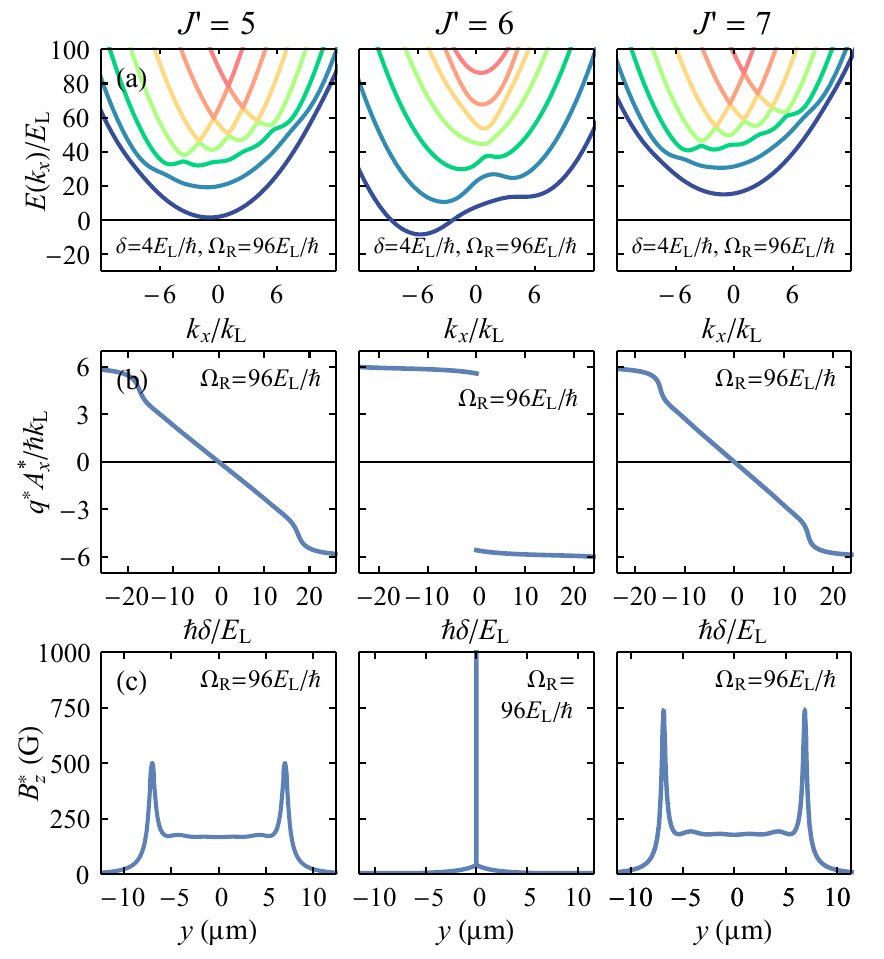}}
\vspace*{-3mm}
\caption{Comparison of results for the different narrow-line transition components \mbox{$J=6 \rightarrow J^\prime=5,6,$} and $7$ (left, middle, and right panels respectively) of the erbium transition, assuming \mbox{$\Omega_{\mathrm{R}} = 96 E_{\mathrm{L}}/\hbar$} in all cases. (a) Energy wavevector dispersion $E(k_x)$ for a two-photon detuning \mbox{$\delta = 4 E_{\mathrm{L}}/\hbar$}. (b) Synthetic vector potential $A^\ast_x$ versus the two-photon detuning $\delta$, and (c) the synthetic magnetic field versus position $y$ for a transverse gradient of the (real) magnetic field of $73.9 \, \unit{G/cm}$ in the case of $J^\prime = 5$ and of $70.3 \, \unit{G/cm}$ in the case of $J^\prime = 7$, for which in both cases $\hbar \omega_{\mathrm{c}} = E_{\mathrm{L}}$ is reached in the center. For both the \mbox{$J=6 \rightarrow J^\prime=5$} and $J^\prime=7$ transitions for the used parameters the lowest energetic dispersion curve has a single minimum, allowing for the synthetization of a -- within the central region \mbox{--} spatially relatively homogeneous synthetic magnetic field. For the case of the \mbox{$J=6 \rightarrow J^\prime=6$} transition the lowest energy dispersion curve for the same value of the Raman coupling has two minima, with the absolute minimum alternating from $k_{x,\mathrm{min}} < 0$ to $k_{x,\mathrm{min}} > 0$ for $\delta>0$ and $\delta<0$ respectively, so that the synthetic vector potential exhibits a step-like behaviour. The resulting expected synthetic magnetic field (shown here for a transverse gradient of the real magnetic field of $70.3 \, \unit{G/cm}$) exhibits a divergence at $y = 0$, as understood from the here discontinuous variation of the vector potential versus $\delta$.}
\label{fig:Comparison}
\end{figure}

We have also investigated the use of the $J = 6 \rightarrow J^\prime = 5$ and $J^\prime = 6$ components of the $4f^{12}6s^2({}^3\textrm{H}_6) \rightarrow 4f^{11}({}^4\textrm{I}^0_{15/2})5d_{5/2}6s^2(15/2, 5/2)_{J^\prime}^0$ erbium transition to implement synthetic magnetic fields. The top panel of Fig.~\ref{fig:Comparison} gives dispersion curves for the $J = 6 \rightarrow J^\prime = 5,6,7$ transitions ((a) - (c)) for \mbox{$\delta = 4 E_{\mathrm{L}}/\hbar$} and \mbox{$\Omega_{\mathrm{R}} = 96 E_{\mathrm{L}}/\hbar$}. The middle and lower panels show the detuning dependence of the  synthetic vector potential and the spatial variation of the synthetic magnetic field respectively. While for the \mbox{$J = 6 \rightarrow J^\prime = 5$} component we expect to reach a spatially quite uniform synthetic magnetic field and obtain $\hbar \omega_{\mathrm{c}} = E_{\mathrm{L}}$ in the center with comparable parameters for the transverse magnetic field gradient, for the \mbox{$J = 6 \rightarrow J^\prime = 6$} case the synthetic field essentially reduces to a single spike in the center. This is understood from the less favorable variation of Clebsch-Gordan coefficients with the Zeeman quantum number, with relatively small couplings near the center of the Zeeman diagram ($|m_F| \approx 0$). Thus, the lowest energetic dispersion curve has two, rather than a single minimum. Tab.~\ref{tab:Gradients} gives a comparison of the required gradients of the real magnetic field to reach a value of the cyclotron frequency of $\hbar \omega_{\mathrm{c}}/E_{\mathrm{L}} = 1$ at $y = 0$ for different values of the two-photon Rabi frequency $\Omega_{\mathrm{R}}$ for both the $J = 6 \rightarrow J^\prime = 5$ and $J = 6 \rightarrow J^\prime = 7$ transitions. As described above, for the lower values of $\Omega_{\mathrm{R}}$, while requiring smaller Raman beam intensities and gradients of the real magnetic field, the spatial homogeneity of the synthetic gauge field reduces.

\section{IV. Laughlin-Gap}
In order to observe the FQHE all atoms subject to the synthetic gauge field have to be in the lowest Landau level (LLL). The following calculations assume a LLL with a filling factor of $\nu = 1/2$ and use theory results from \cite{Regnault2003} in which an ensemble of atoms subject to $s$-wave interactions is considered. In the presence of interactions, the true ground state then becomes a highly correlated Laughlin state. To allow for a selective loading by adiabatic mapping from e.g. an initial Bose-Einstein condensate, the energetic gap to the next excited state, the so-called Laughlin gap $\Delta E_{\mathrm{LG}}$, should be sufficiently large.

Given experimental limits on the experimentally realizable flux of the gauge field, the use of small atom numbers may seem desirable. In the following, we assume that the besides the usual $s$-wave interactions additional dipole-dipole interactions present for the erbium case do not introduce significant modifications to the described picture. We note that in the case of longer-range interactions, such as $1/r^3$ couplings due to dipole-dipole interactions, higher order Haldane pseudopotentials increase in importance. As a consequence the ground state may not be well described by a Laughlin state. However, as discussed for example for the case of bosons with van-der-Waals $1/r^6$ interactions in \cite{Grusdt2013} for large filling fractions, such as $\nu=1/2$, the ground state is still a Laughlin state. Specifically, consider a disk-shaped trapping geometry, with the confinement along the axis of the synthetic magnetic field (i.e. the $z$-axis) being sufficiently strong to restrict the atomic dynamics to the two transverse directions (i.e. in the $x$-$y$ plane).

In the case of $N = 4$ atoms per microtrap the Laughlin gap was estimated to $\Delta E_{\mathrm{LG}} \approx 0.16 g_{\mathrm{int}}$ where \mbox{$g_{\mathrm{int}} = \sqrt{32 \pi} \hbar \omega_{\mathrm{c}} a_{\mathrm{s}} / \ell_z$} is the 2D interaction coefficient, $a_{\mathrm{s}}$ the $s$-wave scattering length and \mbox{$\ell_z = \sqrt{\hbar / m \omega_z}$} the confinement length in $z$-direction, with $\omega_z$ being the corresponding trapping frequency. The disk-shaped configuration can e.g. be realized by the dipole potential induced by a far-detuned one-dimensional standing wave with wavelength  $\lambda_{\mathrm{trap}}$. In this configuration we have \mbox{$\omega_z / (2 \pi)= \sqrt{2 U_0 / m} / \lambda_{\mathrm{trap}}$}, where $U_0$ denotes the trap depth. For \mbox{$\lambda_{\mathrm{trap}} = 1.064 \, \unit{\mu m}$} and a typical trap depth $U_0 = 50 E_{\mathrm{L,trap}}$, with $E_{\mathrm{L,trap}} = h^2 / (2 m \lambda^2_{\mathrm{trap}})$ we arrive at $\ell_z \simeq 64 \, \unit{nm}$ and $\omega_{\mathrm{c}}/(2 \pi) = 14.8 \, \unit{kHz}$. For $\hbar \omega_{\mathrm{c}} = E_{\mathrm{L}}$, as we expect to achieve using parameters described in section III, we arrive at a Laughlin gap of $\Delta E_{\mathrm{LG}} \approx h \cdot 720 \, \unit{Hz}$ for the case of a Raman beams wavelength tuned to near the $J = 6 \rightarrow J^\prime = 7$ transition and a $s$-wave scattering length of $a_s = 200 a_0$ \cite{Frisch2014}, where $a_0$ is Bohr's radius. For larger atom number the predicted size of the Laughlin gap slightly reduces, and in the asymptotic case ($N \gg 1$) reaches $\Delta E_{\mathrm{LG}} \simeq 0.1 g_{int}$, corresponding to $\approx h \cdot 450 \, \unit{Hz}$ for the above parameters. For the corresponding gap sizes adiabatic loading from a Bose-Einstein condensate seems realistic. Also larger atom numbers per trap are experimentally feasible. Here one benefits from the incompressibility of the Laughlin phase, pushing quasi-holes to the outer trap regions. This is a useable configuration when applying spatially resolved detection techniques only monitoring the central trap region.

\section{V. Conclusions}
We have investigated the laser-induced synthetization of gauge fields in the atomic erbium lanthanide system with a ground-state orbital angular momentum $L>0$. A configuration with two counterpropagating oppositely circular polarized Raman beams was shown to be an attractive approach for both on \mbox{$J = 6 \rightarrow J^\prime = 5$} and \mbox{$J^\prime = 7$} narrow-line atomic erbium transitions. In the presence of a transverse gradient of the real magnetic field, strong synthetic magnetic fields with good spatial homogeneity are predicted to be possible, with estimated photon scattering rates roughly two orders of magnitude lower than in implementations with alkali-metal atomic systems. We have moreover estimated the size of the Laughlin gap arising from $s$-wave interactions for typical experimental parameters. Our result suggests that rare-earth atomic systems are attractive candidates for experimental investigations of fractional quantum Hall physics. In view of the large magnetic dipole moments an important topic for future theory work is the investigation of the effect of dipole-dipole interactions (DDI) on the form of the ground state in the presence of the synthetic magnetic field \cite{Lahaye2007,Chomaz2019}.

\section*{Acknowledgments} 
We thank M. Lepers for fruitful discussions. Financial support of the Deutsche Forschungsgemeinschaft (Grant No. CRC-TR 185, Project No. 277625399) and the Cluster of Excellence ML4Q (Project No. EXC 2004/1-390534769) is acknowledged.

\appendix*

\section{APPENDIX}

In Tab.~\ref{tab:CGC} the Clebsch-Gordan coefficients for the here relevant ground state sublevels of the three transitions with $J = 6 \rightarrow J' = 5,6,7$ respectively are listed. Here $m_F$ denotes the ground state sublevel from which a transition to $m_{F'} = m_F + 1$ (with $\Delta m_F = + 1$) or $m_F - 1$ (with $\Delta m_F = - 1$) originates.

\renewcommand{\arraystretch}{2.2} 
\setlength{\tabcolsep}{5.5pt}      
\begin{table}
\begin{adjustbox}{width=\columnwidth,center}
    \begin{tabular}{|c|c|c|c|c|c|c|c|}
    \hline
    $m_F$ & $-6$ & $-4$ & $-2$ & $0$ & $2$ & $4$ & $6$\\ \hline
    \multicolumn{8}{|c|}{$J = 6 \rightarrow J' = 5$}\\ \hline
    $\Delta m_F = + 1$ & $\sqrt{\frac{11}{13}}$ & $\sqrt{\frac{15}{26}}$ & $\sqrt{\frac{14}{39}}$ & $\sqrt{\frac{5}{26}}$ & $\frac{1}{\sqrt{13}}$ & $\frac{1}{\sqrt{78}}$ & \\ \hline
    $\Delta m_F = - 1$ & & $\frac{1}{\sqrt{78}}$ & $\frac{1}{\sqrt{13}}$ & $\sqrt{\frac{5}{26}}$ & $\sqrt{\frac{14}{39}}$ & $\sqrt{\frac{15}{26}}$ & $\sqrt{\frac{11}{13}}$\\ \hline
    \multicolumn{8}{|c|}{$J = 6 \rightarrow J' = 6$}\\ \hline
    $\Delta m_F = + 1$ & $-\frac{1}{\sqrt{7}}$ & $-\sqrt{\frac{5}{14}}$ & $-\sqrt{\frac{10}{21}}$ & $-\frac{1}{\sqrt{2}}$ & $-\sqrt{\frac{3}{7}}$ & $-\sqrt{\frac{11}{42}}$ & \\ \hline
    $\Delta m_F = - 1$ & & $\sqrt{\frac{11}{42}}$ & $\sqrt{\frac{3}{7}}$ & $\frac{1}{\sqrt{2}}$ & $\sqrt{\frac{10}{21}}$ & $\sqrt{\frac{5}{14}}$ & $\frac{1}{\sqrt{7}}$\\ \hline
    \multicolumn{8}{|c|}{$J = 6 \rightarrow J' = 7$}\\ \hline
    $\Delta m_F = + 1$ & $\frac{1}{\sqrt{91}}$ & $\sqrt{\frac{6}{91}}$ & $\sqrt{\frac{15}{91}}$ & $\frac{2}{\sqrt{13}}$ & $3 \sqrt{\frac{5}{91}}$ & $\sqrt{\frac{66}{91}}$ & $1$\\ \hline
    $\Delta m_F = - 1$ & $1$ & $\sqrt{\frac{66}{91}}$ & $3 \sqrt{\frac{5}{91}}$ & $\frac{2}{\sqrt{13}}$ & $\sqrt{\frac{15}{91}}$ & $\sqrt{\frac{6}{91}}$ & $\frac{1}{\sqrt{91}}$\\ \hline
    \end{tabular}
\end{adjustbox}
\caption{Relevant Clebsch-Gordan coefficients for the three transitions $J = 6 \rightarrow J' = 5,6,7$. \label{tab:CGC}}
\end{table}



\begin{thebibliography}{1000}

 \bibitem{Yoshioka2002} See, e.g., D. Yoshioka, \textit{The Quantum Hall Effect} (Springer-Verlag, Berlin, 2002).
 
 \bibitem{Tsui1982} D. C. Tsui, H. L. Stormer, and A. C. Gossard, Phys. Rev. Lett. \textbf{48}, 1559 (1982).
 
 \bibitem{Bretin2004} V. Bretin, S. Stock, Y. Seurin, and J. Dalibard, Phys. Rev. Lett. \textbf{92}, 050403 (2004).
  
 \bibitem{Schweikhard2004} V. Schweikhard, I. Coddington, P. Engels, V. P. Mogendorff, and E. A. Cornell, Phys. Rev. Lett. \textbf{92}, 040404 (2004).
 

 \bibitem{Hauke2012} P. Hauke, O. Tieleman, A. Celi, C. \"Olschl\"ager, J. Simonet, J. Struck, M. Weinberg, P. Windpassinger, K. Sengstock, M. Lewenstein, and A. Eckardt, Phys. Rev. Lett. \textbf{109}, 145301 (2012).

 \bibitem{Lin2009} Y.-J. Lin, R. L. Compton, K. Jimenez-Garcia, J. V. Porto, and I. B. Spielman, Nature \textbf{462}, 628 (2009).

 \bibitem{Aidelsburger2013} M. Aidelsburger, M. Atala, M. Lohse, J. T. Barreiro, B. Paredes, and I. Bloch, Phys. Rev. Lett. \textbf{111}, 185301 (2013).

 \bibitem{Kennedy2013} C. J. Kennedy, G. A. Siviloglou, H. Miyake, W. C. Burton, and W. Ketterle, Phys. Rev. Lett. \textbf{111}, 225301 (2013).

 \bibitem{Abrikosov1957} A. A. Abrikosov, J. Phys. Chem. Solids \textbf{2}, 199 (1957).
 
 \bibitem{Cooper2001} N. R. Cooper, N. K. Wilkin, and J. M. F. Gunn, Phys. Rev. Lett. \textbf{87}, 120405 (2001).
 
 \bibitem{Regnault2003} N. Regnault and T. Jolicoeur, Phys. Rev. Lett. \textbf{91}, 030402 (2003).
  
 \bibitem{Paredes2001} B. Paredes, P. Fedichev, J. I. Cirac, and P. Zoller, Phys. Rev. Lett. \textbf{87}, 010402 (2001).
 
 \bibitem{Juzeliunas2004} G. Juzeli\={u}nas and P. \"Ohberg, Phys. Rev. Lett. \textbf{93}, 033602 (2004).
 
 \bibitem{Ruseckas2005} J. Ruseckas, G. Juzeli\={u}nas, P. \"Ohberg, and M. Fleischhauer, Phys. Rev. Lett \textbf{95}, 010404 (2005).
 
 \bibitem{Juzeliunas2006} G. Juzeli\={u}nas, J. Ruseckas, P. \"Ohberg, and M. Fleischhauer, Phys. Rev. A \textbf{73}, 025602 (2006).
 
 \bibitem{Zhu2006} S.-L. Zhu, H. Fu, C.-J. Wu, S.-C. Zhang, and L.-M. Duan, Phys. Rev. Lett \textbf{97}, 240401 (2006).
 
 \bibitem{Spielman2009} I. B. Spielman, Phys. Rev. A \textbf{79}, 063613 (2009).
 
 \bibitem{Cui2013} X. Cui, B. Lian, T.-L. Ho, B. L. Lev, and H. Zhai, Phys. Rev. A \textbf{88}, 011601(R) (2013).
 
 
 \bibitem{Lu2011} M. Lu, N.Q. Burdick, S.-H. Youn, and B. L. Lev, Phys. Rev. Lett. \textbf{107}, 190401 (2011).
  
 \bibitem{Aikawa2012} K. Aikawa, A. Frisch, M. Mark, S. Baier, A. Rietzler, R. Grimm, and F. Ferlaino, Phys. Rev. Lett. \textbf{108}, 210401 (2012).
 
 \bibitem{Aikawa2014} K. Aikawa, A. Frisch, M. Mark, S. Baier, R. Grimm, and F. Ferlaino, Phys. Rev. Lett. \textbf{112}, 010404 (2014).
 
 \bibitem{Ulitzsch2017} J. Ulitzsch, D. Babik, R. Roell, and M. Weitz, Phys. Rev. A \textbf{95}, 043614 (2017).
 
 \bibitem{Kalganova2015} E. Kalganova, G. Vishnyakova, A. Golovisin, D. Tregubov, D. Sukachev, S. Fedorov, K. Khabarova, A. Akimov, N. Kolachevsky, and V. Sorokin, J. Phys.: Conf. Ser. \textbf{635}, 092117 (2015).

 
 \bibitem{Dalibard2015} J. Dalibard, in \textit{Proceedings of the International School of Physics "Enrico Fermi" on Quantum Matter at Ultralow Temperatures}, edited by M. Inguscio, W. Ketterle, S. Stringari, and G. Roati (Società Italiana di Fisica, Italy, 2015).
 
 
 \bibitem{Kadau2016} H. Kadau, M. Schmitt, M. Wenzel, C. Wink, T. Maier, I. Ferrier-Barbut and T. Pfau, Nature \textbf{530}, 194-197 (2016).
 
 \bibitem{Baier2018} S. Baier, D. Petter, J. H. Becher, A. Patscheider, G. Natale, L. Chomaz, M. J. Mark, and F. Ferlaino, Phys. Rev. Lett. \textbf{121}, 093602 (2018).
 
 \bibitem{Ban2005} H. Ban, M. Jacka, J. Hanssen, J. Reader, and J. McClelland, Opt. Express \textbf{13}, 3185 (2005).
 
 \bibitem{Lepers2014} M. Lepers, J. F. Wyart, and O. Dulieu, Phys. Rev. A \textbf{89}, 022505 (2014).
 
 \bibitem{Martin2017} A. M. Martin, N. G. Marchant, D. H. J. O'Dell, and N. G. Parker, J. Phys.: Condens. Matter \textbf{29}, 103004 (2017).
  
 \bibitem{Grusdt2013} F. Grusdt and M. Fleischhauer, Phys. Rev. A \textbf{87}, 043628 (2013).
 
 \bibitem{Frisch2014} A. Frisch, \textit{Dipolar Quantum Gases of Erbium}, Dissertation, University of Innsbruck, (2014). 
 
 \bibitem{Lahaye2007} T. Lahaye, T. Koch, B. Fr\"olich, M. Fattori, J. Metz, A. Griesmaier, S. Giovanazzi, and T. Pfau, Nature \textbf{448}, 672-675 (2007).
 
 \bibitem{Chomaz2019} L. Chomaz, D. Petter, P. Ilzh\"{o}fer, G. Natale, A. Trautmann, C. Politi, G. Durastante, R. M. W. van Bijnen, A. Patscheider, M. Sohmen, M. J. Mark and F. Ferlaino, Phys. Rev. X \textbf{9}, 021012 (2019).
 

%
%
%
%
%
%
%
%
%
%
%
%
%
%
%
%
%
%
%
%
%
%
%
%
%
%
%
%
%

\end{thebibliography}


\end{document}